\documentclass[epj]{webofc}
\usepackage[varg]{txfonts}   
\usepackage{subfigure}
\usepackage{amsmath}
\usepackage{multirow}
\usepackage{color}
\usepackage{lineno}
\begin{document}
%
\selectlanguage{english}
\title{Prospects of LHC Higgs Physics at the end of Run III}
%
%

\author{Xin Chen\inst{1,2,3}{\fnsep\thanks{\email{xin.chen@cern.ch}}, On behalf of the ATLAS and CMS Collaborations}
}

\institute{Department of Physics, Tsinghua University, Beijing 100084, China
\and Collaborative Innovation Center of Quantum Matter, Beijing 100084, China
\and Center for High Energy Physics, Peking University, Beijing 100084, China
}

\abstract{%
The document is prepared for the LCWS2016 conference proceedings. The expected status of Higgs physics at the end of Run-3 is presented. The current Run-2 status is briefly reviewed, and the expected Higgs reach after the HL-LHC period is also summarized for some channels.
}
\maketitle
\section{Introduction}
\label{intro}

The Large Hadron Collider (LHC) is taking data at $\sqrt{s}=13$ TeV centre of mass energy. At the end of Run-2 (2018), an integrated luminosity of 150 $fb^{-1}$ is expected, and at the end of Run-3, 300 $fb^{-1}$ is expected, after which is the HL-LHC period until about 3 $ab^{-1}$ of data is collected. There are two major detector upgrades happening after Run-2 and Run-3. Phase-I upgrade (2019-2020) prepares for an instantaneous luminosity of 2-$3\times 10^{34}\text{cm}^{-2}\text{s}^{-1}$. The production of detector parts for Phase-I has already started. The Phase-II upgrade (2024-2026) prepares for an instantaneous luminosity of 5-$7.5\times 10^{34}\text{cm}^{-2}\text{s}^{-1}$, and copes with an average pileup events of 140-200 per $p$-$p$ collision. Phase-II is currently in the design and R\&D stage. The major detector upgrades for the CMS and ATLAS experiments are outlined in Tab.~\ref{tab:tab1}:
\begin{table}[h]
\caption{ Major expected upgrades to the CMS and ATLAS detectors after the Phase-I and Phase-II upgrades. }
\label{tab:tab1}
\centering 
\begin{tabular}{|c|c|c|}
\hline 
 & CMS & ATLAS \\ \hline
 Tracking & Extended to $\left| \eta \right|<3.8$ & Extended to $\left| \eta \right|<4.0$. All silicon \\ \hline
 Calorimeter & \multicolumn{2}{c|}{Update all readout electronics. Timing in EM endcap (to reject pileup)} \\ \hline
Trigger & \multicolumn{2}{c|}{Tracking added at L1, larger bandwith, finer granularity} \\ \hline
\multirow{2}{*}{Muon} & New chamber to complete $1.6<\left| \eta \right|<2.4$, & New endcap wheel to reject \\ 
 & muon tagger up to  $\left| \eta \right|<3.0$ & fake L1 muons (Phase-I) \\ \hline
\end{tabular}
\end{table}

\begin{figure}[h]
\centering
\includegraphics[width=6cm,clip]{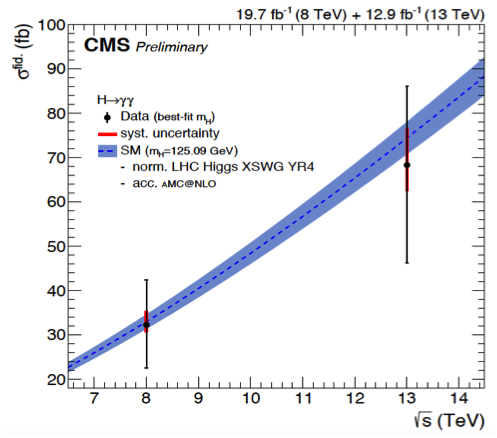}\hspace{5mm}
\includegraphics[width=7.0cm,clip]{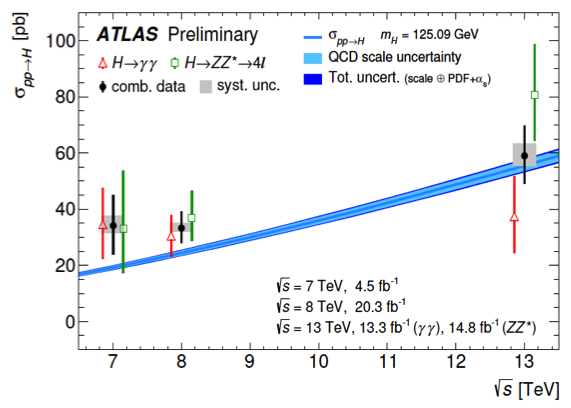}
\caption{ The measured cross sections of $H\to\gamma\gamma$ and $H\to ZZ^*\to 4l$ with initial Run-2 $\sqrt{s}=13$ TeV data at CMS (left) \cite{ref4} and ATLAS (right) \cite{ref3}. }
\label{fig:fig1}
\end{figure}

The fiducial cross section measurement results with initial Run-2 data at $\sqrt{s}=13$ TeV data are given in Tab.~\ref{tab:tab2}. The Higgs cross sections roughly scale with the center of mass energy, as displayed in Fig.~\ref{fig:fig1} \cite{ref1,ref2,ref3,ref4,ref5}.
\begin{table}[h]
\caption{ The predicted and measured Higgs production fiducial cross sections with initial Run-2 $\sqrt{s}=13$ TeV data at ATLAS and CMS \cite{ref1,ref2,ref4,ref5}. The luminosity used by ATLAS is 13.3 $fb^{-1}$ (14.8 $fb^{-1}$) for $H\to\gamma\gamma$ ($H\to ZZ^*\to 4l$), and 12.9 $fb^{-1}$ for CMS.}
\label{tab:tab2}
\centering 
\begin{tabular}{|c|c|c|c|c|}
\hline 
 & \multicolumn{2}{c|}{ $H\to\gamma\gamma$} & \multicolumn{2}{c|}{ $H\to ZZ^*\to 4l$ } \\ \cline{2-5}
 & $\sigma_{\text{fid}}$ (fb) & SM pred. (fb) & $\sigma_{\text{fid}}$ (fb) & SM pred. (fb) \\ \hline
 ATLAS & $43.2\pm 14.9(\text{stat})\pm 4.9(\text{sys})$ & $62.8_{-4.4}^{+3.4}(\text{N3LO})$ & $4.54_{-0.90}^{+1.02}$ & $3.07_{-0.25}^{+0.21}$ \\ \hline
 CMS & $69_{-22}^{+16}(\text{stat})_{-6}^{+8}(\text{sys})$ & $73.8\pm 3.8$ & $2.29_{-0.64}^{+0.74}(\text{stat})_{-0.23}^{+0.30}(\text{sys})$ & $2.53\pm 0.13$ \\ \hline
\end{tabular}
\end{table}

With about 10 times of the existing Run-1 data set, and larger Higgs cross sections, we are able to (1) precisely measure Higgs production and decay rates and couplings, (2) test the Higgs sector and probe for Beyond Standard Model (BSM) such as Minimal Supersymmetric Standard Model (MSSM), the double Higgs production rate (order of few percent effects on Higgs couplings in most models), (3) search for rare/new/invisible decay modes, (4) use Effective Field Theory for the Higgs tensor structure study. Current theory errors on the Higgs gluon-gluon-Fusion (ggF) and Vector-Boson-Fusion (VBF) production cross sections are about 3-4\% and 0.5\% for the scale, 3\% and 2\% for PDF+$\alpha_s$, respectively \cite{ref6}. The theory errors on the Higgs Branching Ratios (BR) are typically at 3-5\%. Many projection results  have particular assumptions made. The systematics can change and the analysis method can also improve. Projections for 300 $fb^{-1}$ and 3 $ab^{-1}$ (HL-LHC period) luminosities are given for some channels in this document.
Physics projections for HL-LHC Higgs measurements are usually done in two ways: 
\begin{itemize}
\item Parametrized detector performance. Event-generator level particles are smeared with detector performance parametrized from full simulation and reconstruction of upgraded HL-LHC detectors. Effects of pile-up are included for either $5\times 10^{34} \text{cm}^{-2}\text{s}^{-1}$ (average 140 pile-up events which is the old default) or $7\times 10^{34} \text{cm}^{-2}\text{s}^{-1}$ (average 200 pile-up events, current new setup). Analysis is mostly based on the existing 8 TeV results with simple re-optimization for higher luminosity.
\item Extrapolation of Run-1 or Run-2 results. This means scaling signal and background to higher luminosities, correcting for different center-of-mass energy, assuming unchanged analysis (not re-optimized for higher luminosity) and the same detector performance as in Run-1/2 (some used corrections based on studies in the first approach).
\end{itemize}
For CMS, sometimes the projections are made with different systematics assumptions as listed in Tab.~\ref{tab:tab1b}, which may appear in different projection plots and tables.
\begin{table}[h]
\caption{ The four different projection schemes at CMS with different assumptions of the systematics \cite{ref12}. }
\label{tab:tab1b}
\centering 
\begin{tabular}{l|c|c|c|c}
\hline \hline 
 & systematics & exp. sys. & theo. sys. & high PU \\ 
 & unchanged & scaled $1/\sqrt{\mathcal{L}}$ & scaled 1/2 & effects \\
ECFA16 S1 &   {\color{green} \checkmark} & {\color{red} $\times$}  & {\color{red} $\times$}  & {\color{red} $\times$} \\
ECFA16 S1+ &   {\color{green} \checkmark} & {\color{red} $\times$}  & {\color{red} $\times$}  & {\color{green} $\checkmark$} \\
ECFA16 S2 &   {\color{red} $\times$} & {\color{green} \checkmark}  & {\color{green} \checkmark}  & {\color{red} $\times$} \\
ECFA16 S2+ &   {\color{red} $\times$} & {\color{green} \checkmark}  & {\color{green} \checkmark}  & {\color{green} \checkmark} \\
\hline \hline
\end{tabular}
\end{table}

\section{Projections for the signal strength and coupling factors}
\label{coupling}

\begin{figure}[h]
\centering
\includegraphics[width=4.5cm,clip]{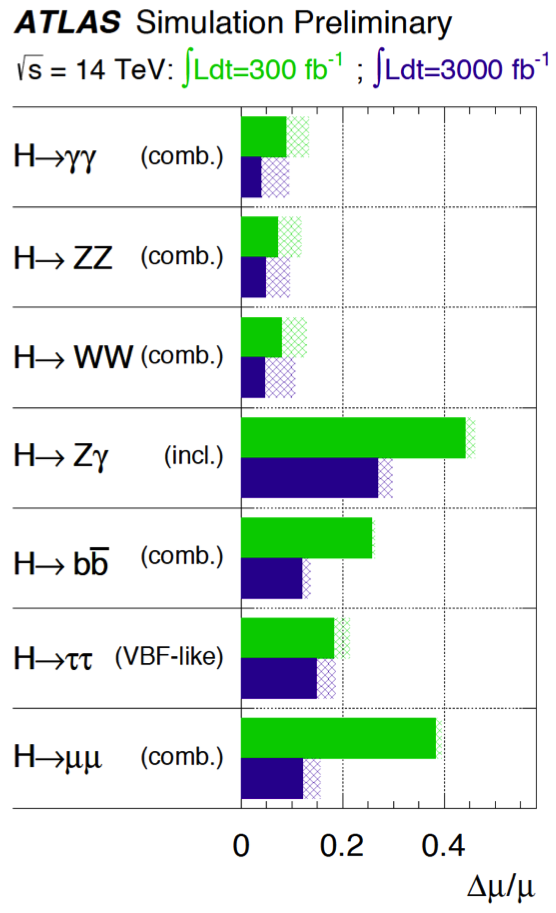}\hspace{10mm}
\includegraphics[width=4.9cm,clip]{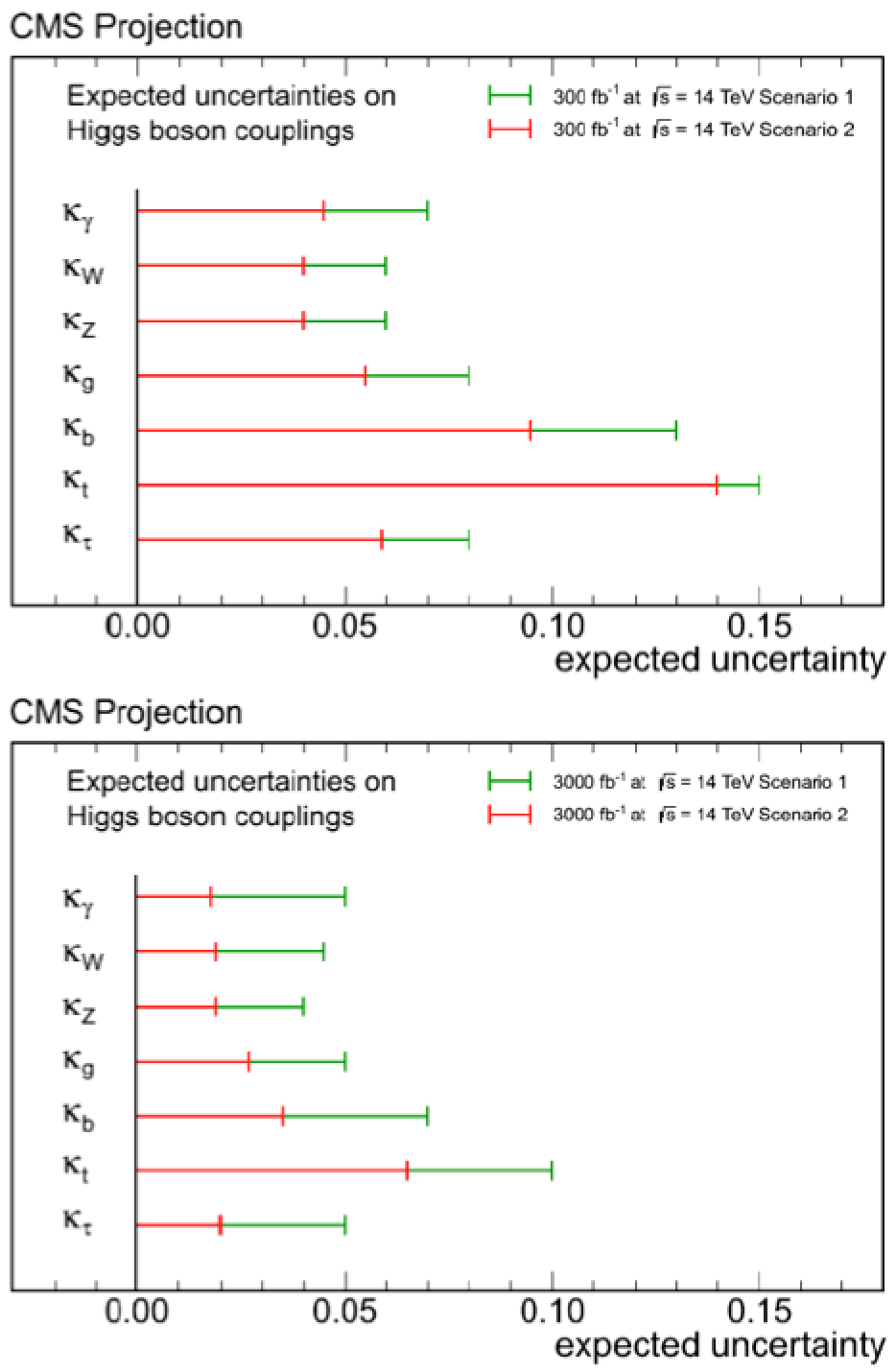}
\caption{ The projected uncertainty on the signal strength in different decays for ATLAS (left) \cite{ref7}, and on the Higgs coupling to different particles for CMS (right) \cite{ref8}, with 300 $fb^{-1}$ and to 3 $ab^{-1}$ of data. }
\label{fig:fig2}
\end{figure}

\begin{figure}[h]
\centering
\includegraphics[width=5.6cm,clip]{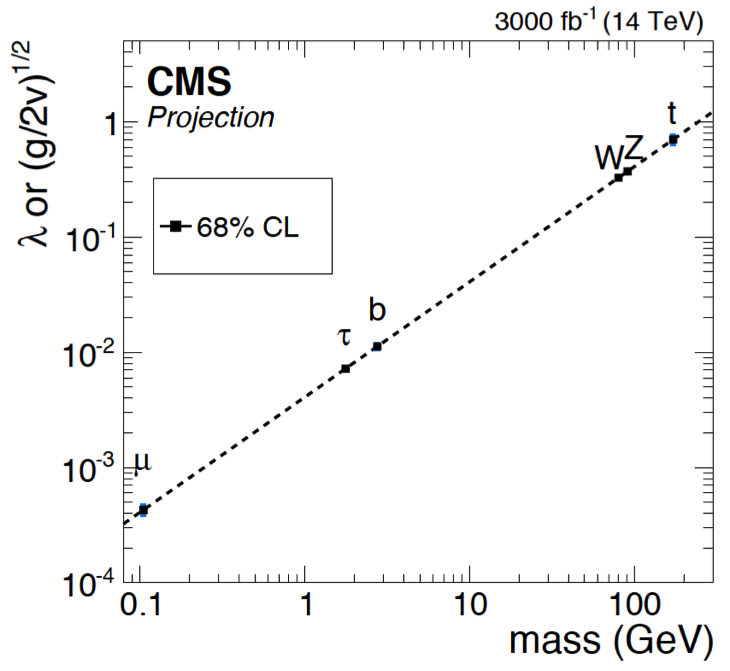}\hspace{10mm}
\includegraphics[width=5.0cm,clip]{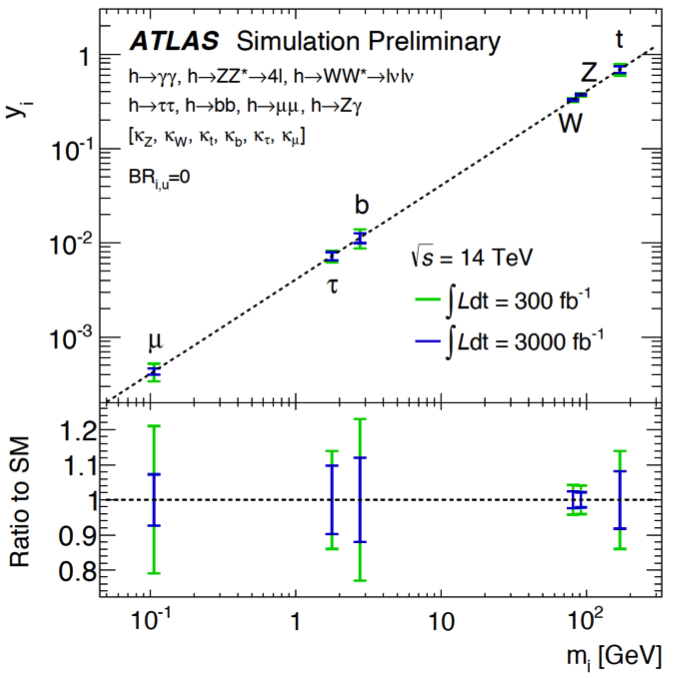}
\caption{ The reduced couplings of Higgs with different particles as a function of Higgs mass at CMS (left) \cite{ref9} and ATLAS (right) \cite{ref7}. }
\label{fig:fig3}
\end{figure}

The projected uncertainty on the Higgs signal strength in different decays for ATLAS, and on the Higgs coupling to different particles for CMS, are shown in Fig.~\ref{fig:fig2} \cite{ref7,ref8}. With larger statistics, the reduced coupling scale factors can be calculated to test the Higgs coupling dependence on the mass, as shown in Fig.~\ref{fig:fig3} \cite{ref7,ref9}. The estimated precisions of such couplings with 3 $ab^{-1}$ of data at ATLAS are about 3\% for the $W/Z$ boson, 7\% for the muon, and 8-12\% for the $t/b/\tau$ fermions. 

\section{Projections for $H\to ZZ^*\to 4l$}
\label{4lep}

\begin{figure}[h]
\centering
\includegraphics[width=5.cm,clip]{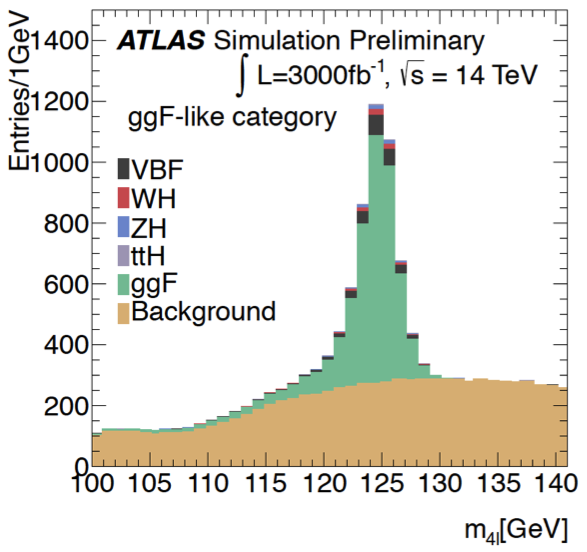}\hspace{10mm}
\includegraphics[width=6.5cm,clip]{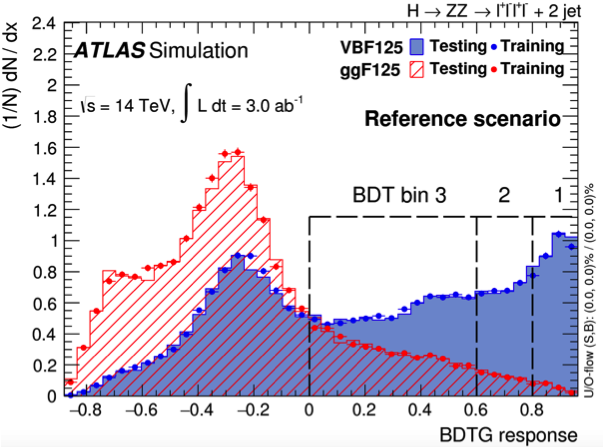}
\caption{ The Higgs mass distribution with different Higgs production contributions overlaid with 5 $ab^{-1}$ (left) \cite{ref10}, and the BDT distribution in the VBF region to separate the ggF and VBF productions, in the $H\to ZZ^*\to 4l$ channel \cite{ref11} at ATLAS. }
\label{fig:fig4}
\end{figure}

\begin{figure}[h]
\centering
\includegraphics[width=5.6cm,clip]{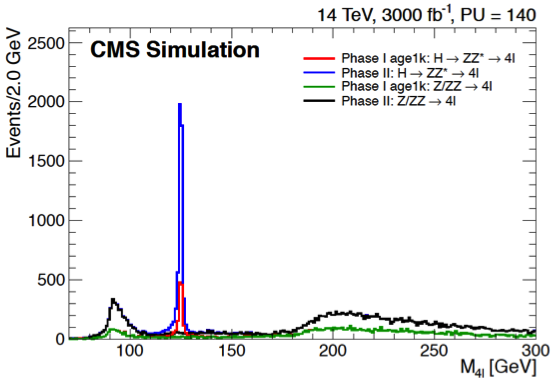}
\includegraphics[width=7.2cm,clip]{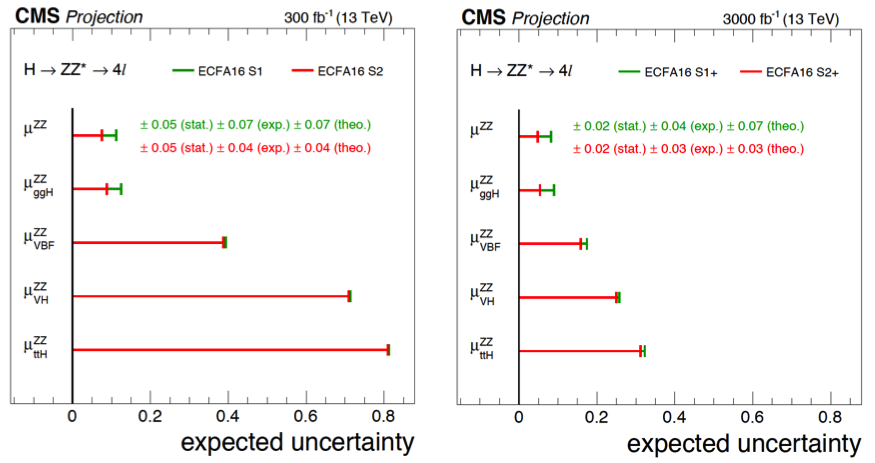}
\caption{ The Higgs mass distribution (left) and the signal strengths in different production modes with 300 $fb^{-1}$ (middle) and 3 $ab^{-1}$ (right) of data in the $H\to ZZ^*\to 4l$ channel at CMS \cite{ref8,ref12}. }
\label{fig:fig5}
\end{figure}

\begin{figure}[h]
\centering
\includegraphics[width=11.cm,clip]{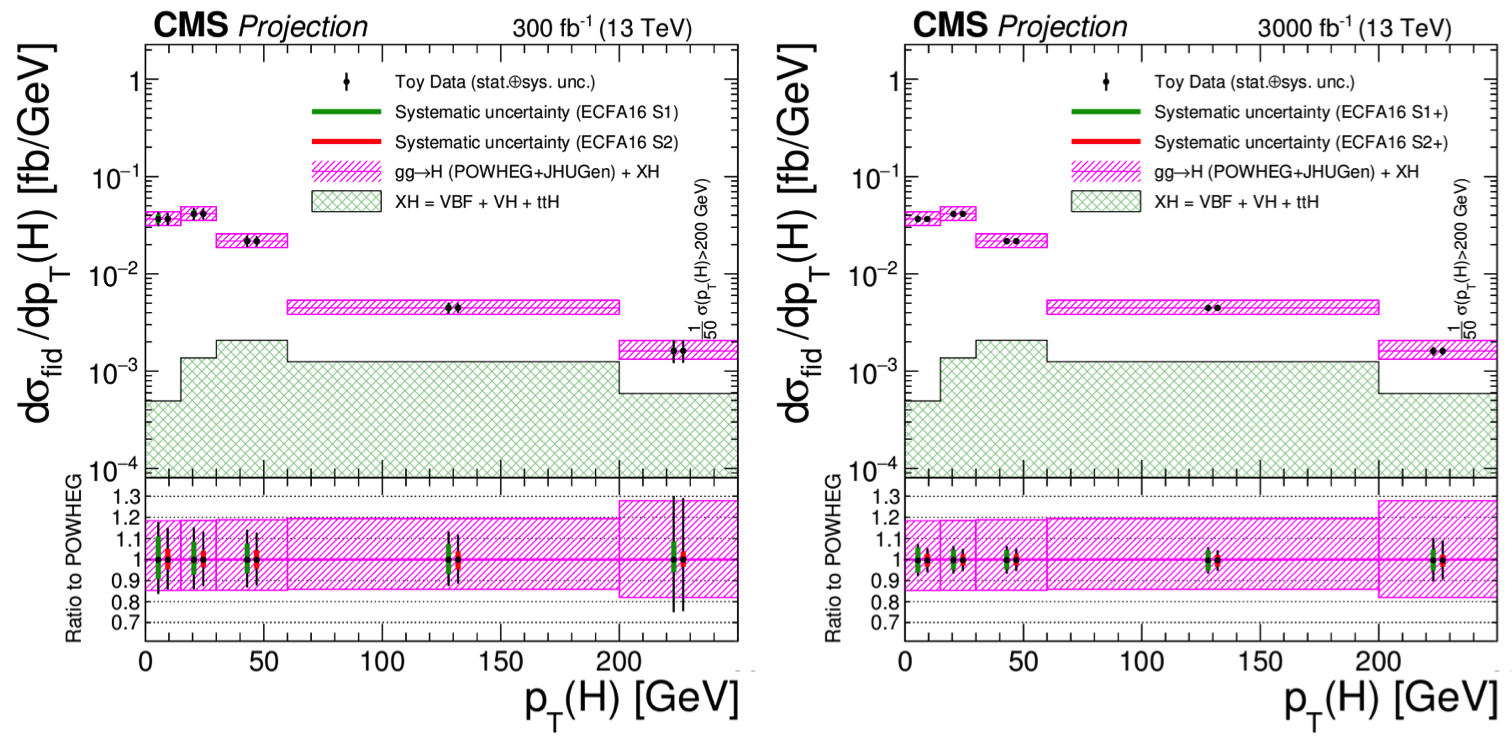}
\caption{ The differential distributions of the Higgs $p_T$ after fiducial cuts with 300 $fb^{-1}$ (left) and 3 $ab^{-1}$ (right) of data in the $H\to ZZ^*\to 4l$ channel at CMS \cite{ref5,ref12}. }
\label{fig:fig6}
\end{figure}

The $H\to ZZ^*\to 4l$ channel is relatively clean. With more data such as 3 $ab^{-1}$, the signal can be divided into VBF, VH, $t\bar{t}H$ and ggH categories, as shown in Fig.~\ref{fig:fig4} \cite{ref10,ref11}. With 300 $fb^{-1}$ of data, the combined uncertainty on the signal strength, $\Delta\mu/\mu=0.125$, can be achieved. The VBF region is almost background free, and a multivariate analysis, in this case the Boosted Decision Trees (BDT), can be used to further separate the ggF and VBF productions. The BDT distribution is also shown in Fig.~\ref{fig:fig4}, assuming an extended tracking coverage is implemented. With an average pileup events of 140 (200), $\Delta\mu/\mu=0.17$ (0.18) and a signal significance of $7.7\sigma$ ($7.2\sigma$) can be reached. The Higgs mass distribution and the signal strengths in different production modes at CMS are shown in Fig.~\ref{fig:fig5} \cite{ref8,ref12}. A $\Delta\mu/\mu$ of 5-10\% can be achieved on the combined production of $H\to ZZ^*\to 4l$ (dominated by ggH) with 3 $ab^{-1}$ at CMS. With abundant signal events, differential distributions such as the Higgs $p_T$ can be also measured as shown in Fig.~\ref{fig:fig6} \cite{ref5,ref12}. It is seen that going from 300 $fb^{-1}$ to 3 $ab^{-1}$, the statistical and some systematic errors are greatly reduced (less than the theory errors at NLO). 

\begin{figure}[h]
\centering
\includegraphics[width=6.cm,clip]{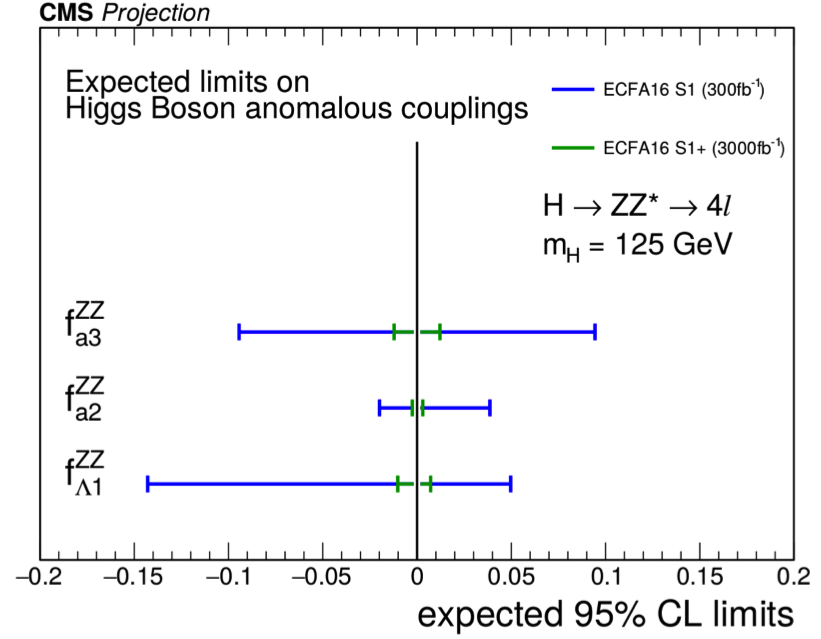}
\caption{ The expected limits on the anomalous Higgs couplings with 300 $fb^{-1}$ and 3 $ab^{-1}$ of data in the $H\to ZZ^*\to 4l$ channel at CMS \cite{ref12}. }
\label{fig:fig7}
\end{figure}

The $H\to ZZ^*\to 4l$ channel can be also used to probe the anomalous tensor couplings in the following effective Lagrangian assuming spin-0 state of the Higgs:
\begin{equation}
\mathcal{L}_{HVV} \sim a_1 \frac{m_Z^2}{2} HZ^\mu Z_\mu - \frac{\kappa_1}{\Lambda_1^2} m_Z^2 H Z_\mu \square Z^\mu - \frac{1}{2} a_2 H Z^{\mu\nu}Z_{\mu\nu} - \frac{1}{2} a_3 HZ^{\mu\nu}\tilde{Z}_{\mu\nu}, 
\label{eq:eq1}
\end{equation}
where the terms other than $a_1$ are BSM dimension-6 operators. Using the phase space distributions of the four leptons in the final state, limits on the coefficients from BSM terms can be set as in Fig.~\ref{fig:fig7} \cite{ref12}, in which the $f_{a_i}$ is defined as
\begin{equation}
f_{a_i} = \frac{\left| a_i \right|^2 \sigma_i}{\sum_j \left| a_j \right|^2 \sigma_j},
\label{eq:eq2}
\end{equation}
where $\sigma_i$ is the cross section from the $i'$th term in Eq.~\ref{eq:eq1}. As the measurement is statistically limited, a big improvement in the sensitivity to the anomalous couplings with 3 $ab^{-1}$ of data is expected.

\section{Projections for $H\to\gamma\gamma$}
\label{diphoton}

\begin{figure}[h]
\centering
\includegraphics[width=5.cm,clip]{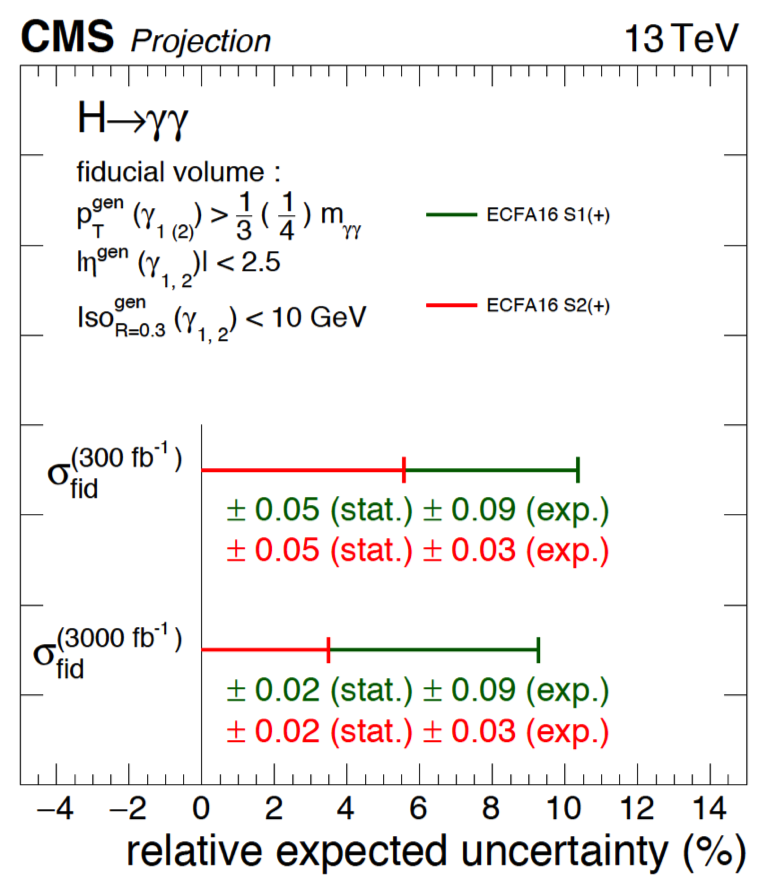}\hspace{10mm}
\includegraphics[width=5.2cm,clip]{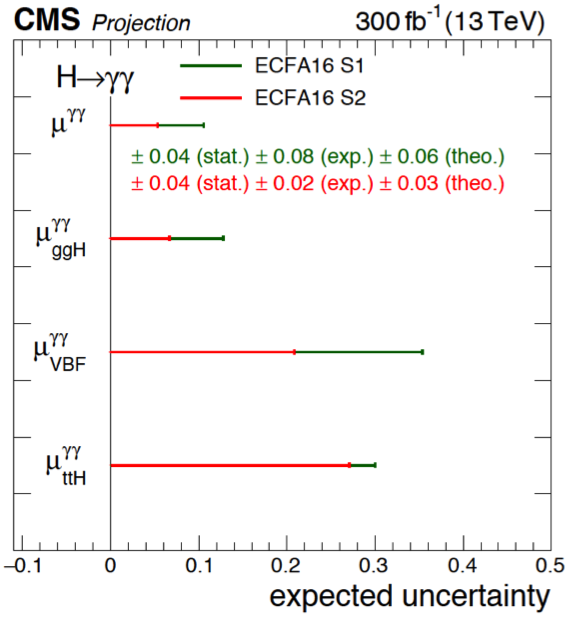}
\caption{ The expected uncertainties for the Higgs fiducial cross section (left), and for the signal strength of different production modes (right) in the $H\to\gamma\gamma$ decay at CMS \cite{ref12}. }
\label{fig:fig8}
\end{figure}

The expected uncertainties for the Higgs fiducial cross section, and for the signal strength of different production modes in the $H\to\gamma\gamma$ decay are shown for CMS in Fig.~\ref{fig:fig8} \cite{ref12}. The fiducial cuts are defined at the generator level. The beam spot uncertainty in the beam direction is simulated to be about 5 cm with degraded vertex efficiency. The photon identification efficiency is degraded by 2.3\% (10\%) for the endcap+barrel (endcap+endcap) photons.

\section{Projections for $H\to Z\gamma$}
\label{Zgam}

\begin{figure}[h]
\centering
\includegraphics[width=5.5cm,clip]{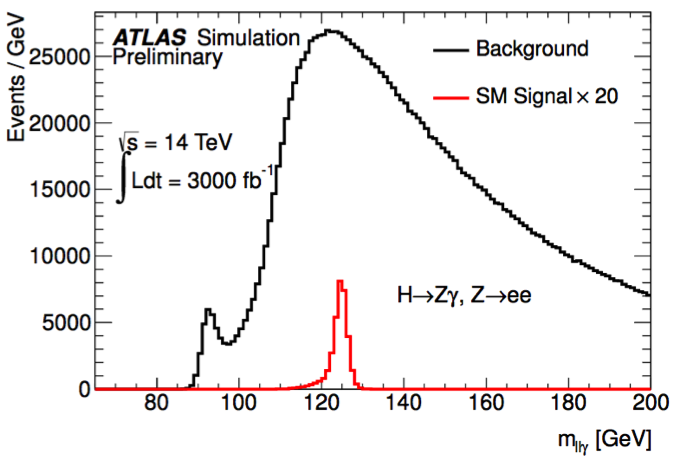}\hspace{10mm}
\includegraphics[width=5.5cm,clip]{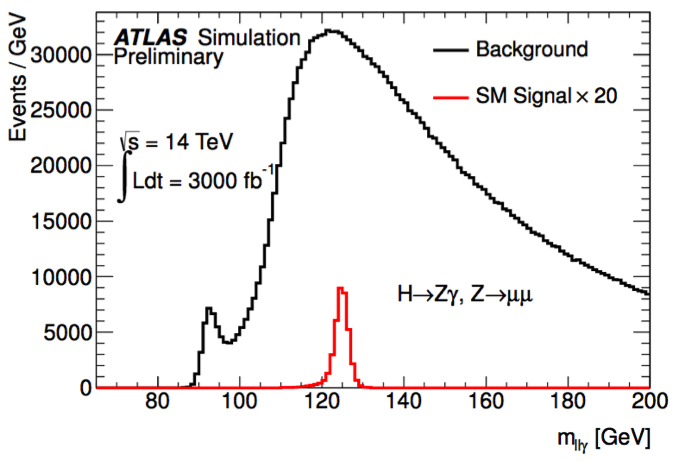}
\caption{ The $Z\gamma$ invariant mass distributions with $Z\to ee$ (left) and $Z\to\mu\mu$ (right) with 3 $ab^{-1}$ of data at ATLAS \cite{ref13}. }
\label{fig:fig9}
\end{figure}

The $H\to Z\gamma$ channel is very challenging due to the high level of SM background. With 300 $fb^{-1}$ (3 $ab^{-1}$) of data, the expected signal significance is $2.3\sigma$ ($3.9\sigma$). However, this channel may be sensitive to BSM through the loops. The $Z\gamma$ invariant mass distributions with $Z\to ee$ and $Z\to\mu\mu$ are shown if Fig.~\ref{fig:fig9} for ATLAS \cite{ref13}.

\section{Projections for $H\to J/\psi\gamma$}
\label{Jgam}

\begin{figure}[h]
\centering
\includegraphics[width=5.5cm,clip]{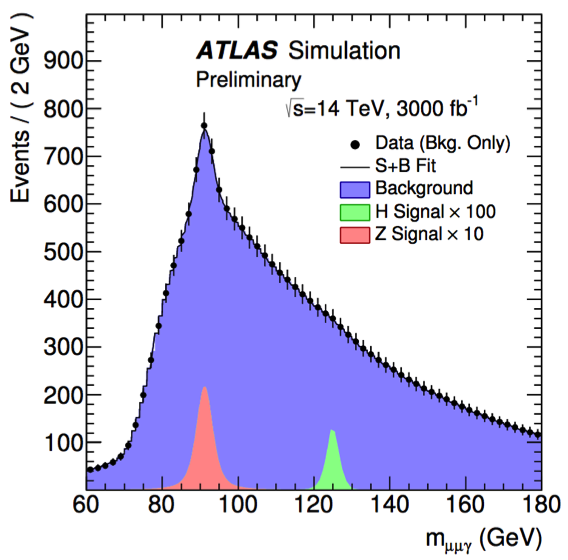}
\caption{ The expected $\mu^+\mu^-\gamma$ mass spectrum in the $H\to J/\psi\gamma$ channel with 3 $ab^{-1}$ of data at ATLAS \cite{ref14}. }
\label{fig:fig9b}
\end{figure}

The $H\to J/\psi\gamma$ is sensitive to the Higgs coupling to charmed quarks. Using Run-1 data and the $J/\psi\to\mu^+\mu^-$ decay, an upper limit on the BR($H\to J/\psi\gamma$) has been set at $1.5\times 10^{-3}$, while the SM prediction gives BR($H\to J/\psi\gamma$)=$(2.9\pm 0.2)\times 10^{-6}$. The $\mu^+\mu^-\gamma$ mass spectrum with 3 $ab^{-1}$ of data is shown in Fig.~\ref{fig:fig9b} \cite{ref14}. With 3 $ab^{-1}$, about 3 signal events are expected out of 1700 background events. Expected limits with Multi-Variate-Analysis (MVA) with no background systematics are BR($H\to J/\psi\gamma$)<$44_{-12}^{+19}\times 10^{-6}$, and $\sigma(gg\to H)\times$BR($H\to J/\psi\gamma$)<$3.1_{-1.3}^{+0.9}$fb. Although the sensitivity is not great, BSM physics can potentially enhance the rate and give an observable signal.

\section{Projections for the Higgs width}
\label{width}

\begin{figure}[h]
\centering
\includegraphics[width=3.4cm,clip]{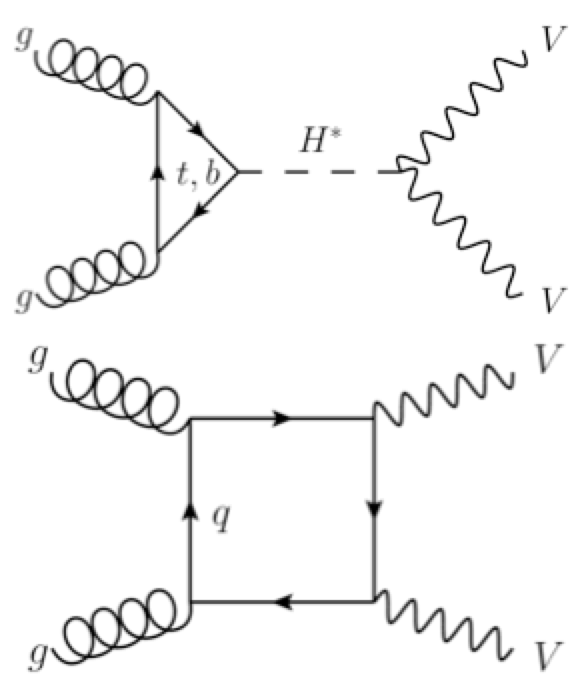}\hspace{3mm}
\includegraphics[width=4.3cm,clip]{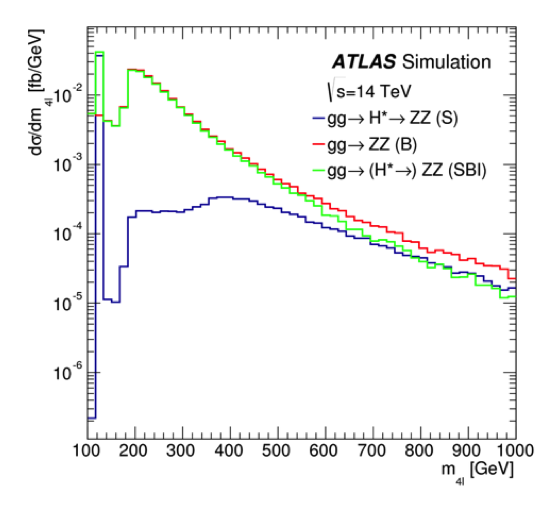}\hspace{3mm}
\includegraphics[width=4.4cm,clip]{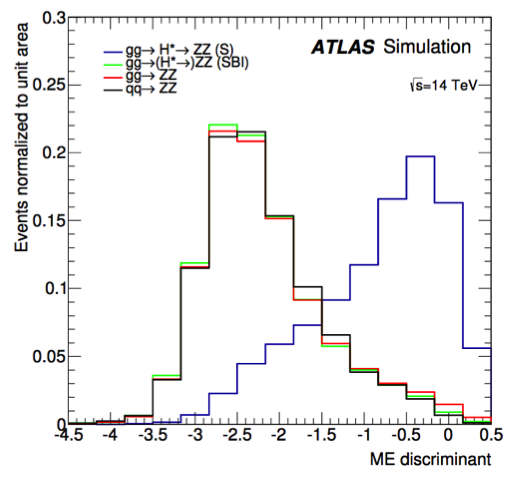}
\caption{ The Feynman diagrams for the $gg\to VV$ process (left),  the distributions of the 4-lepton mass (middle) and the ME (right) in the $ZZ\to 4l$ final state at ATLAS \cite{ref15}. }
\label{fig:fig10}
\end{figure}

The Higgs width can be measured by the interference between the triangle and box diagrams in the $gg\to VV$ process, as illustrated in Fig.~\ref{fig:fig10}. With the $H\to ZZ\to 4l$ final state, the 4-lepton mass shape and the Matrix Element (ME) for each event can be used to discriminate the signal from background. Fig.~\ref{fig:fig10} also shows the 4-lepton mass and ME distributions \cite{ref15}. When combined with the on-shell measurements of the Higgs, the Higgs width can be estimated. In Run-1 with the three decays of $ZZ\to 4l$, $ZZ\to 2l2\nu$ and $WW\to e\nu\mu\nu$, upper limits of $\mu_{\text{offshell}}<6.2$ (8.1), and $\Gamma_H/\Gamma_H^{SM}<5.5$ (8.0) for the observation (expectation), were obtained. With 3 $ab^{-1}$, the precision can be improved to $\mu_{\text{offshell}}=1.00_{-0.50}^{0.43}$ and $\Gamma_H=4.2_{-2.1}^{+1.5}$ MeV (statistical + systematic errors).

\section{Projections for $H\to\mu\mu$}
\label{Hmumu}

\begin{figure}[h]
\centering
\includegraphics[width=4.0cm,clip]{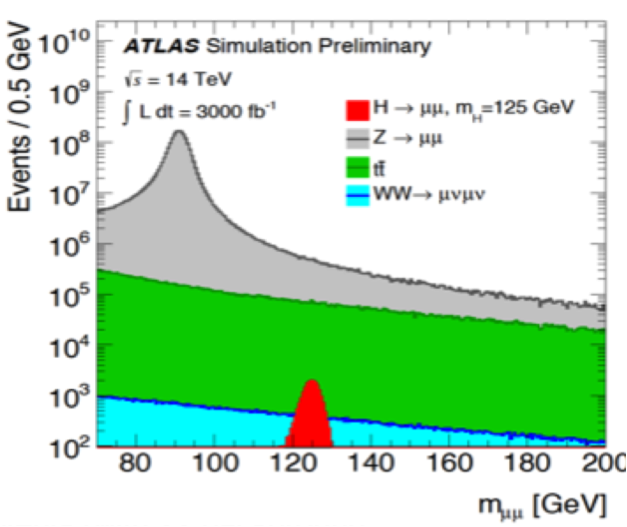}\hspace{1mm}
\includegraphics[width=4.7cm,clip]{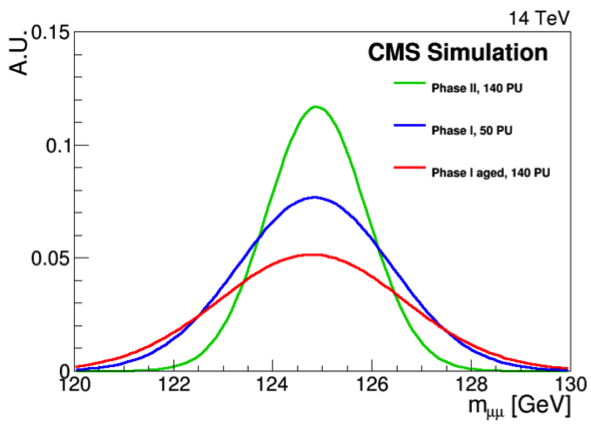}\hspace{1mm}
\includegraphics[width=4.6cm,clip]{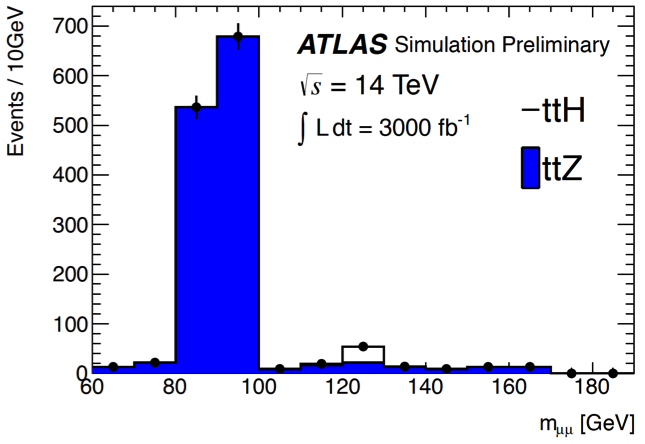}
\caption{ The dimuon mass spectrum with 3 $ab^{-1}$ of data for ATLAS (left), the dimuon mass resolution for different detector and pileup conditions for CMS (middle), and the dimuon mass in the $t\bar{t}H$ associated production for ATLAS (right) \cite{ref8,ref10}. }
\label{fig:fig11}
\end{figure}

The $H\to\mu\mu$ channel is very challenging because of the low BR. In Fig.~\ref{fig:fig11} \cite{ref8,ref10}, the dimuon mass spectrum with 3 $ab^{-1}$ of data, the dimuon mass resolution for different detector and pileup conditions, and the dimuon mass in the $t\bar{t}H$ associated production are given. The sharpness of the dimuon mass is crucial for both ATLAS and CMS. The $H\mu\mu$ coupling can be measured to a precision of 8-20\% with 3 $ab^{-1}$. In the $t\bar{t}H$, $H\to\mu\mu$ channel, about 33 signal events can be selectred out of about 22 background events, and with 300 $fb^{-1}$, $\Delta\mu/\mu=46\%$ and a signal significance of $2.3\sigma$ can be reached.

\section{Projections for the double Higgs production}
\label{HH}

\begin{figure}[h]
\centering
\includegraphics[width=4.4cm,clip]{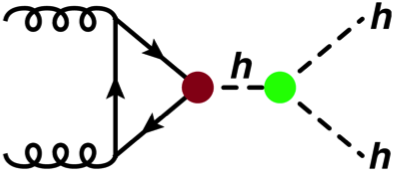}\hspace{5mm}
\includegraphics[width=3.6cm,clip]{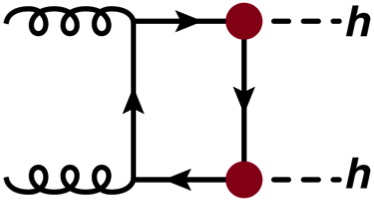}\hspace{5mm}
\includegraphics[width=4.4cm,clip]{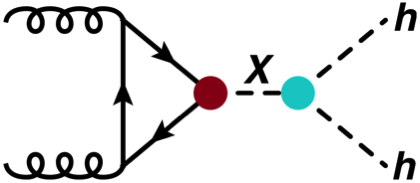}
\caption{ The SM (left, middle) and BSM (right) diagrams that contribute to the double Higgs production at the LHC \cite{ref12}. }
\label{fig:fig12}
\end{figure}

\begin{figure}[h]
\centering
\includegraphics[width=5.8cm,clip]{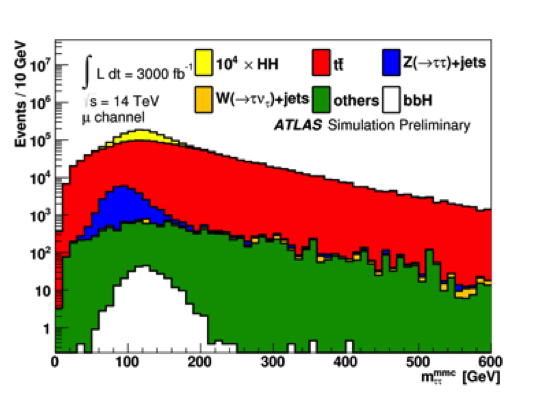}\hspace{5mm}
\includegraphics[width=4.0cm,clip]{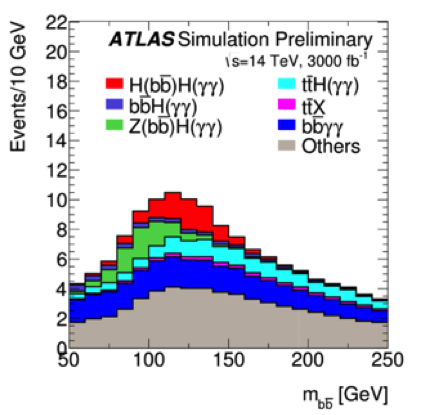}\hspace{5mm}
\includegraphics[width=5.8cm,clip]{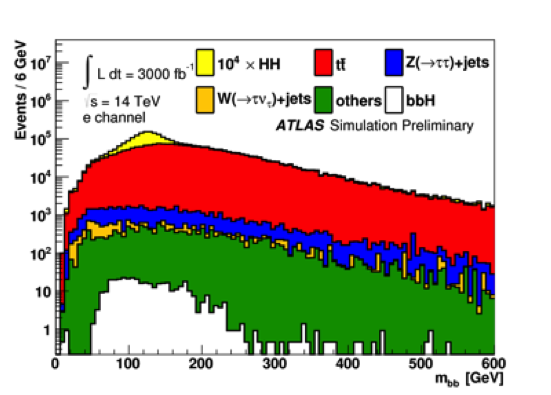}\hspace{5mm}
\includegraphics[width=4.0cm,clip]{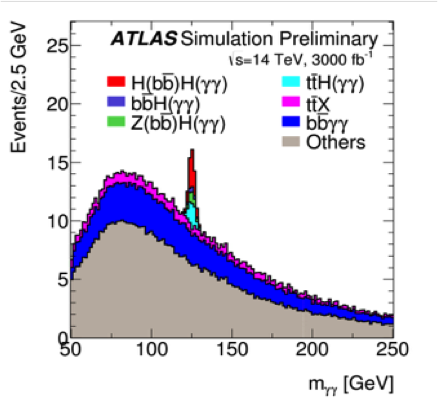}
\caption{ The invariant mass of $\gamma\gamma$ and $b\bar{b}$ in the $HH\to b\bar{b}\tau\tau$ channel (left column), and of the $b\bar{b}$ and $\gamma\gamma$ pairs in the $HH\to b\bar{b}\gamma\gamma$ channel (right column) with 3 $ab^{-1}$ of data at ATLAS \cite{ref16,ref17,ref18,ref19}. }
\label{fig:fig13}
\end{figure}

\begin{figure}[h]
\centering
\includegraphics[width=4.5cm,clip]{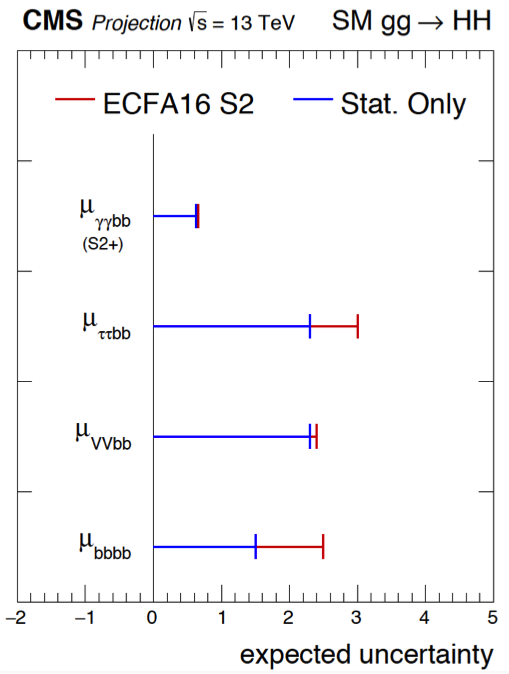}
\caption{ The expected uncertainties on the signal strength in different double Higgs decay modes with 3 $ab^{-1}$ of data at CMS \cite{ref12}. }
\label{fig:fig14}
\end{figure}

The Higgs self-coupling is measured with the double Higgs production. The relevant Feynman diagrams are shown in Fig.~\ref{fig:fig12}.
For the non-resonant SM double Higgs production, the triangle and box diagrams have negative interference, and the NNLO+NNLL prediction of the cross section is $\sigma=33.45$ fb at 13 TeV. The invariant mass of $\gamma\gamma$ and $b\bar{b}$ in the $HH\to b\bar{b}\tau\tau$ channel, and of the $b\bar{b}$ and $\gamma\gamma$ in the $HH\to b\bar{b}\gamma\gamma$ channel with 3 $ab^{-1}$ of data at ATLAS are shown in Fig.~\ref{fig:fig13} \cite{ref16,ref17,ref18,ref19}. The signal significances for $HH\to b\bar{b}\gamma\gamma$, $HH\to b\bar{b}\tau\tau$ and $t\bar{t}HH\to t\bar{t}b\bar{b}b\bar{b}$ are $1.3\sigma$, $0.6\sigma$ and $0.35\sigma$ respectively. The 95\% CL interval for the Higgs quartic coupling constant is $-1.3<\lambda/\lambda_{SM}<8.7$ ($-7.4<\lambda/\lambda_{SM}<14$) in the $HH\to b\bar{b}\gamma\gamma$ ($HH\to b\bar{b}b\bar{b}$) channel. The expected uncertainties on the signal strength in different double Higgs decay modes from CMS are shown in Fig.~\ref{fig:fig14} \cite{ref12}, and the upper limits on the signal strength are also listed in Tab.~\ref{tab:tab3} \cite{ref12}. Sensitivity for the SM double Higgs production is still quite limited; anyway several BSM models can be excluded.
%
\begin{table}[h]
\caption{ The expected upper limits on the signal strength of the double Higgs production signal in different decay modes with 3 $ab^{-1}$ of data at CMS \cite{ref12}. }
\label{tab:tab3}
\centering 
\begin{tabular}{|l|c|c|c|c|c|c|c|c|c|}
\hline
Channel & \multicolumn{3}{c|}{Median expected} & \multicolumn{3}{c|}{Z-value} & \multicolumn{3}{c|}{Uncertainty} \\
 & \multicolumn{3}{c|}{limits in $\mu_r$} & \multicolumn{3}{c|}{} & \multicolumn{3}{c|}{as fraction of $\mu_r=1$} \\
 & \multicolumn{2}{c|}{ECFA16} & Stat. & \multicolumn{2}{c|}{ECFA16} & Stat. & \multicolumn{2}{c|}{ECFA16} & Stat. \\
 & S1 & S2 & Only & S1 & S2 & Only & S1 & S2 & Only \\ \hline
 $gg\to HH\to \gamma\gamma bb$ (S1+/S2+) & 1.3 & 1.3 & 1.3 & 1.6 & 1.6 & 1.6 & 0.64 & 0.64 & 0.64 \\
 $gg\to HH\to \tau\tau bb$ & 7.4 & 5.2 & 3.9 & 0.28 & 0.39 & 0.53 & 3.7 & 2.6 & 1.9 \\
 $gg\to HH\to VVbb$ &  & 4.8 & 4.6 &  & 0.45 & 0.47 &  & 2.4 & 2.3 \\
 $gg\to HH\to bbbb$ &  & 7.0 & 2.9 &  & 0.39 & 0.67 &  & 2.5 & 1.5 \\ \hline
\end{tabular}
\end{table}

\section{Projections for MSSM $H\to\tau\tau$}
\label{MSSM_Htt}

\begin{figure}[h]
\centering
\includegraphics[width=6.5cm,clip]{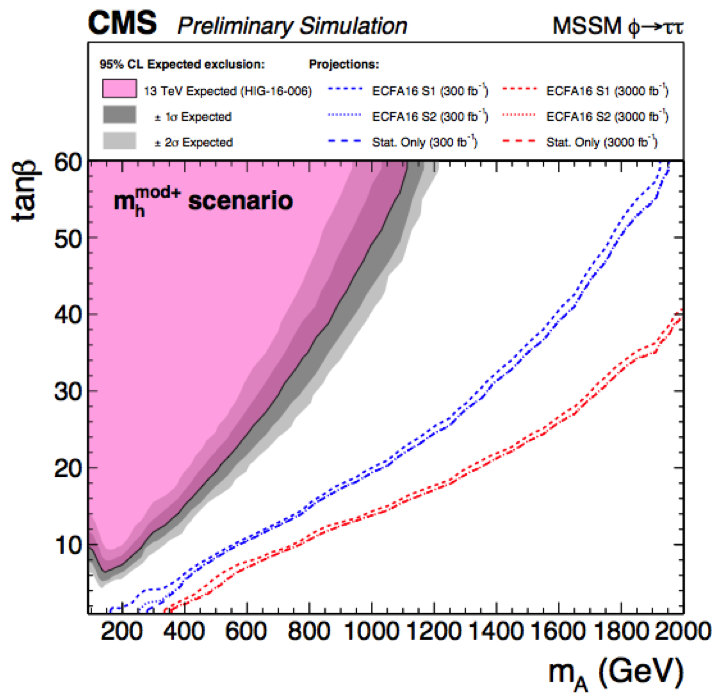}
\caption{ The expected exclusion region in the $m_A$-$\tan\beta$ plane in the $m_h^{\text{mod}+}$ scenario for the MSSM $\phi\to\tau\tau$ decay at CMS \cite{ref12}. }
\label{fig:fig15}
\end{figure}

In MSSM, there are five Higgs scalars, namely $h$, $A$, $H$ and $H^\pm$. To be compatible with a 125 GeV Higgs, usually two scenarios are considered:
\begin{itemize}
\item hMSSM scenario: the measured value of 125 GeV can be used to predict masses and decay branching ratios of the other Higgs bosons.
\item $m_h^{\text{mod}+}$ scenario: the lightest CP-even Higgs is assigned to be the 125 GeV boson.
\end{itemize}
The expected exclusion region in the $m_A$-$\tan\beta$ plane in the $m_h^{\text{mod}+}$ scenario for the $\phi\to\tau\tau$ decay at CMS is shown in Fig.~\ref{fig:fig15} \cite{ref12}.

\section{Projections for VBF $H\to\text{invisible}$}
\label{H_invisible}

\begin{figure}[h]
\centering
\includegraphics[width=6.5cm,clip]{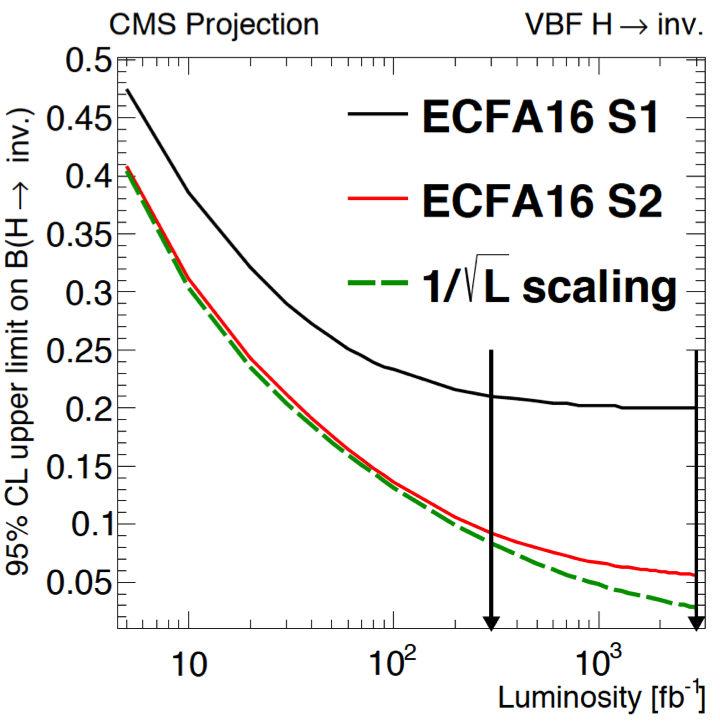}
\caption{ The projected limits on $H\to\text{invisible}$ BR with the VBF production as a function of integrated luminosity from CMS \cite{ref12}. }
\label{fig:fig16}
\end{figure}

The Higgs can decay to dark matter particles such as the neutralinos leading to large Missing $E_T$ (MET) events. To select these events, MET triggers are required, and the VBF region with large $m_{jj}$ and $\Delta\eta_{jj}$ cuts are investigated. Dedicated $Z\to ll$ and $W\to l\nu$ control regions are available to check the background from single vector boson processes. The VBF channel is found to be more powerful than the VH, $H\to\text{invisible}$ channel. With Run-1 data, the limit of BR$(H\to\text{invisible})<0.28$ (0.31) was obtained for the observation (expectation) from ATLAS \cite{ref20}. The projected limits with the VBF production as a function of integrated luminosity from CMS are shown in Fig.~\ref{fig:fig16} \cite{ref12}. 

\section{Conclusion}
\label{conclusion}
With a successful startup of LHC Run-2, both CMS and ATLAS are fully engaged in the 13 TeV centre of mass energy running. To exploit at best the LHC upgrade (Run III and  HL-LHC), both experiments have to cope with the challenges of increased pileup rates with dedicated upgrades to the detectors. At the end of Run 3 with 300 $fb^{-1}$, the Higgs couplings can be measured to 10-20\%. At the end of HL-LHC with 3 $ab^{-1}$ of data, the precision on the couplings can improve to about 5-10\%, and statistically limited measurements, like the differential Higgs distributions and test for Higgs tensor couplings, can be significantly improved. Rare Higgs decays can be probed as well. For example, the $H\to\mu\mu$ channel can reach $<20\%$ precision, and the sensitivity for the anomalous $H\to J/\psi\gamma$ decay at $10\times$SM level can be achieved. The Higgs self-coupling via double Higgs productions can be searched for, although is still quite challenging. Indirect search for BSM through the Higgs width, and direct search for new signatures such as $H\to\text{invisible}$ and the MSSM Higgs will make tighter tests of the theory. Higgs precision physics is one of the main motivations and design considerations for the HL-LHC and experiments programs. 

%
%
%

\end{document}